\documentclass[aps,pre,reprint,groupedaddress]{revtex4-1}

\usepackage[pdftex]{graphicx} 

\newcommand{\syn}{W_{ij}}

\begin{document}

% Use the \preprint command to place your local institutional report
% number in the upper righthand corner of the title page in preprint mode.
% Multiple \preprint commands are allowed.
% Use the 'preprintnumbers' class option to override journal defaults
% to display numbers if necessary
%\preprint{}

%Title of paper
\title{Critical neural networks with short and long term plasticity}

% repeat the \author .. \affiliation  etc. as needed
% \email, \thanks, \homepage, \altaffiliation all apply to the current
% author. Explanatory text should go in the []'s, actual e-mail
% address or url should go in the {}'s for \email and \homepage.
% Please use the appropriate macro foreach each type of information

% \affiliation command applies to all authors since the last
% \affiliation command. The \affiliation command should follow the
% other information
% \affiliation can be followed by \email, \homepage, \thanks as well.
%\email[]{Your e-mail address}
%\homepage[]{Your web page}
%\thanks{}
%\altaffiliation{}

\author{L. Michiels van Kessenich}
\email[]{laurensm@ethz.ch}
\affiliation{Computational Physics for Engineering Materials, IfB, ETH Z\"urich}
\author{ M. Lukovi\'{c}}
%\email[]{lukovicm@ethz.ch}
\affiliation{Computational Physics for Engineering Materials, IfB, ETH Z\"urich}
\author{ L. de Arcangelis}
%\email[]{lucilla.dearcangelis@unicampania.it}
\affiliation{Dept. of Industrial and Information Engineering, University of Campania "Luigi Vanvitelli", Aversa (CE), Italy}
%\affil[3]{INFN sez. Naples, Gr. Coll. Salerno (Italy)}
\author{ H. J. Herrmann}
\altaffiliation{Departamento de Fisica, Universidade Federal do Cear\'{a}, 60451-970 Fortaleza, Cear\'{a}, Brasil}
%\email[]{hans@ifb.baug.ethz.ch}
\affiliation{Computational Physics for Engineering Materials, IfB, ETH Z\"urich}

%\author[1,*]{L. Michiels van Kessenich}
%\author[1]{ M. Lukovi\'{c}}
%\author[2,3]{ L. de Arcangelis}
%\author[1,4]{ H. J. Herrmann}
%\affil[1]{Computational Physics for Engineering Materials, IfB, ETH Z\"urich}
%\affil[2]{Dept. of Industrial and Information Engineering, University of Campania "Luigi Vanvitelli", Aversa (CE), Italy}
%\affil[3]{INFN sez. Naples, Gr. Coll. Salerno (Italy)}
%\affil[4]{Departamento de Fisica, Universidade Federal do Cear\'{a}, 60451-970 Fortaleza, Cear\'{a}, Brasil}
%\affil[*]{laurensm@ethz.ch}

%Collaboration name if desired (requires use of superscriptaddress
%option in \documentclass). \noaffiliation is required (may also be
%used with the \author command).
%\collaboration can be followed by \email, \homepage, \thanks as well.
%\collaboration{}
%\noaffiliation

\date{\today}

\begin{abstract}
In recent years self organised critical neuronal models have provided insights regarding the origin of the experimentally observed avalanching behaviour of neuronal systems. It has been shown that dynamical synapses, as a form of short-term plasticity, can cause critical neuronal dynamics. Whereas long-term plasticity, such as hebbian or activity dependent plasticity, have a crucial role in shaping the network structure and endowing neural systems with learning abilities. In this work we provide a model which combines both plasticity mechanisms, acting on two different time-scales. The measured avalanche statistics are compatible with experimental results for both the avalanche size and duration distribution with biologically observed percentages of inhibitory neurons. The time-series of neuronal activity exhibits temporal bursts leading to 1/f decay in the power spectrum. The presence of long-term plasticity gives the system the ability to learn binary rules such as XOR, providing the foundation of future research on more complicated tasks such as pattern recognition.
\end{abstract}

% insert suggested PACS numbers in braces on next line
% insert suggested keywords - APS authors don't need to do this
%\keywords{}

%\maketitle must follow title, authors, abstract, \pacs, and \keywords
\maketitle

% body of paper here - Use proper section commands
% References should be done using the \cite, \ref, and \label commands
\section{Introduction}
% Put \label in argument of \section for cross-referencing
%\section{\label{}}
%\subsection{}
%\subsubsection{}

Scale-invariant neuronal dynamics, in the form of avalanches, have been reported by a multitude of experiments measuring spontaneous neuronal activity in \textit{vitro} and in \textit{vivo} \cite{plenz2014criticality,beggsPlenz2003, beggs2004neuronal,plenz2007organizing, mazzoni2007dynamics,petermann2009spontaneous, gireesh2008neuronal, massobrio2015criticality, de2014criticality}. These observations suggest that neuronal systems operate near a critical point. The measured size and duration distributions follow a power-law behaviour with exponents $1.5$ and $2.0$ respectively, which characterize the mean-field self-organised branching process \cite{zapperi1995}. The experimentally observed scale-invariant dynamics has first been numerically reproduced by neuronal models closely related to self-organised criticality (SOC) \cite{lucillaPRL,lucillaBrainCrit,cocchi2017criticality,arcanglis2012dragon,miranda1991self}. SOC provides a mechanism by which the critical point is the attractor of the systems dynamics \cite{pruessner}. It therefore presents an elegant explanation to why these scaling exponents are observed consistently in both {\it in vitro} and {\it in vivo} experiments, ranging from dissociated neurons \cite{pasquale2008self} to MEG measurements on human patients \cite{shriki2013}. A neuronal system is subject to plastic adaptation which follows the principles of Hebbian plasticity \cite{hebb} or activity dependent plasticity, i.e. synapses between neurons whose activity is correlated are strengthened and inactive synapses are weakened. This is a form of long-term plasticity which is believed to be the basis of learning mechanisms\cite{martin2000synaptic,caporale2008spike}. By this activity dependent adaptation the synaptic structure of the network is modified in an attempt to solve problems or classification tasks. Conversely, short-term plasticity originates from mechanisms which operate on shorter time-scales. One of these mechanisms is synaptic depression or fatigue \cite{simons2006synapticdepression}. Repeated activations of neurons lead to the depletion of vesicles which contain the neurotransmitter available at the synapses \cite{ikeda2009vesicles}. The inclusion of synaptic fatigue in critical neuronal dynamics was first proposed by \textit{Levina et al.} \cite{levina,levina2009phase,de2015can}.\\

Here we present a model which includes both short- and long-term plasticity. In addition to plastic adaptation other neurobiological mechanisms influence the dynamics. These include inhibitory neurons and the neuronal refractory time.  An action potential originating from an inhibitory neuron causes the release of inhibitory neurotransmitters, such as GABA \cite{watanabe2002gaba}. This hyper-polarizes the post-synaptic neurons and reduces their firing probability. Regarding the neuronal dynamics, inhibitory neurons can be seen as sinks for the activity, since the activation of an inhibitory neuron leads to the reduction of activity in connected neurons. Moreover, the firing rate of real neurons is limited by the refractory period, i.e., the brief period after the generation of an action potential during which a second action potential cannot be triggered \cite{nicholls2001neuron}.\\

The scale-invariant behaviour of SOC models relies heavily on conservative dynamics \cite{bonachela2009self}. More precisely, one distinguishes between \textit{bulk} and \textit{boundary} dissipation \cite{pruessner}. In traditional SOC models, such as the Bak-Tang-Wiesenfeld-model \cite{BTW}, the \textit{bulk} dynamics is energy conserving whereas the boundary dissipation scales $\propto \sqrt{N}$ and is negligible in the thermodynamic limit $N\rightarrow\infty$. If a system has non-conservative interactions between the individual units it suffers from \textit{bulk}-dissipation. This implies that the dissipation does not vanish in the thermodynamic limit. It is well known that conservation plays a crucial role in SOC models, yet it is worth mentioning that several models with non-conservative dynamics do show apparent scale-invariant behaviour\cite{pruessner}. These systems, such as the OFC earthquake model \cite{ofc1992} or Drossel Schwabel forest-fire model\cite{drossel1992forest}, require a loading mechanism to compensate the dissipative elements. Ultimately, these models rely on the careful tuning of the recharging rate to observe scale invariant dynamics. \textit{Bonachela et al.} \cite{bonachela2009self,bonachela2010self} provided a detailed study of scale-invariant dynamics observed in non-conservative SOC models. The model presented here features non-conserving neuronal dynamics and also relies on a tuning parameter. In this way scale-invariant behaviour is obtained even in the presence of dissipation.\\

The article starts with a presentation of the model and the description of how plastic adaptation operating on two different time scales is implemented. Then we present results regarding the avalanche statistics and power spectrum and compare with experimental findings. In the final section the learning capacities of the model are examined by training the system to learn the XOR rule.

\section{Neuronal model}

Consider a directed neuronal network consisting of $N$ neurons connected by synapses. Each neuron is characterized by a membrane potential $v_i$. To model both long- and short-term plasticity each synapse is represented by the two variables $W_{ij}$ and $w_{ij}$. The short-term synaptic strength $w_{ij}$ and the long-term synaptic strength $W_{ij}$. The initial value for the potentials $v_i$ and the short-term synaptic strengths $w_{ij}$ are not important since the system will evolve towards a steady state, independent of the initial condition. $W_{ij}$ are chosen homogeneously in the interval $[0,W_{max}]$ where $W_{max}$ is a parameter of the model. Other distributions can be chosen for $W_{ij}$ but as we will show the relevant parameter is the average long-term synaptic strength $\langle W\rangle$. If the potential in a neuron surpasses a threshold, $v_i\geq v_c$, it causes the generation of an action potential which travels along the axon towards the synapses. The release of neurotransmitter then causes a change in the potential of the connected post-synaptic neurons $j$ according to:

\begin{equation}
\label{eq:dynSynFiring}
v_j(t+1)=v_j(t)\pm v_i(t)uw_{ij}(t),
\end{equation}
\begin{equation}
\label{eq:dynSyn}
w_{ij}(t+1)=w_{ij}(t)(1-u),
\end{equation}
\begin{equation}
\label{eq:potReset}
v_{i}\rightarrow 0 .
\end{equation}

\begin{figure}
    \includegraphics[width=0.85\columnwidth]{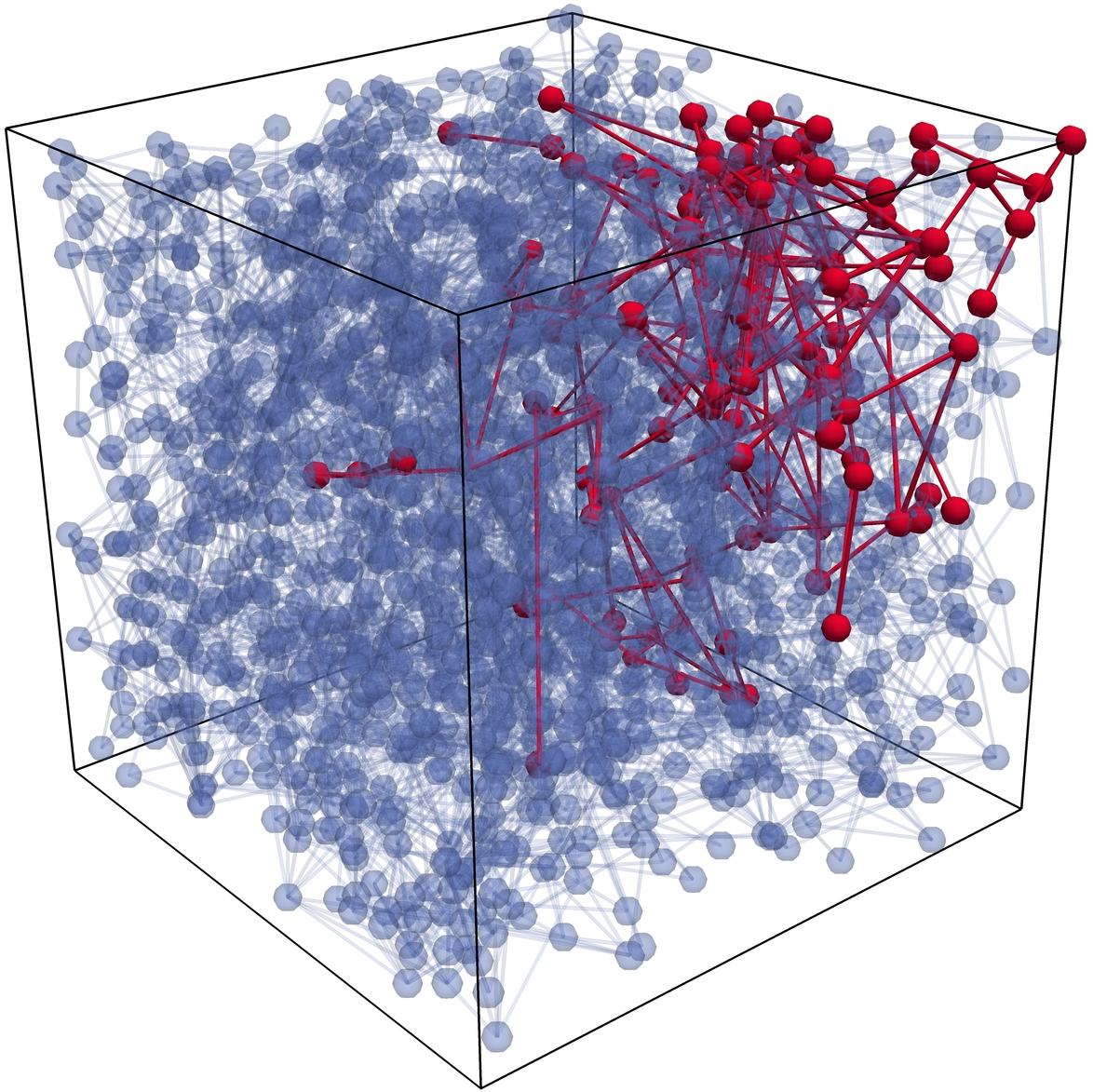}
    \caption{\label{fig:exampleConfig}An avalanche on a scale-free network. Neurons and synapses which participated to the avalanche are shown in red. Inactive neurons and synapses are blue.}
\end{figure}

\noindent Where the $+$ and $-$ stand for excitatory and inhibitory signals respectively. We set a certain percentage $p_{in}$ of the neurons to be inhibitory which are chosen randomly. After a neuron fires it enters a refractory state for $t_r$ time steps. During this period it does not respond to stimulations from other neurons and will not produce further action potentials. $w_{ij}$ represents the available neurotransmitter at the synapse $ij$ and $u$ is the fraction of neurotransmitter released when the synapse activates. The vesicles at a synapse which are ready for immediate release upon an incoming action potential are called the  readily releasable pool (RRP)\cite{kaeser2017readily,rosenmund1996definition}. It has been shown that the RRP at a synapse is of the order of five percent of the total available neurotransmitter\cite{rizzoli2005synpool,ikeda2009vesicles}. We therefore set, $u=0.05$. When the potentials of all neurons are below $v_c$, activity is triggered by adding a small stimulation $\delta v=0.1$ to random neurons until one of them reaches the threshold. If the action potential causes a post-synaptic neuron to surpass $v_c$ neuronal activity propagates until all neurons have potentials $v_i<v_c$. The size $s$ of an avalanche is given by the total number of active neurons and its duration is the number of time-steps the avalanche lasts. Since during the propagation of an avalanche the available neurotransmitter decreases according to equation (\ref{eq:dynSyn}), a recovery mechanism is needed to sustain future activity. Therefore, after each avalanche the available neurotransmitter recovers as $w_{ij}= w_{ij}+\syn$. The short-term synaptic weight $w_{ij}$ determines the strength of the stimulation received by post-synaptic neurons according to equation (\ref{eq:dynSynFiring}). The numerical value for $w_{ij}$ will be several orders of magnitude larger than $W_{ij}$ since it increases by repeated additions of $\syn$. $w_{ij}$ therefore fluctuates on short time-scales and $\syn$ determines around which value $w_{ij}$ fluctuates so that a higher $\syn$ will lead to a higher average $w_{ij}$. In this way, both long- and short-term plasticity are implemented. We call $w_{ij}$ the "short-term synaptic strength" and $\syn$ the "long-term synaptic strength". A key parameter is the average long-term synaptic strength $\langle W\rangle=\sum_{ij}\syn/N_s$ where $N_s$ is the number of synapses.  Depending on $\langle W\rangle$ the system will exhibit more or less neuronal activity. The value for $\langle W\rangle$ is set by choosing the initial values $\syn$ homogeneously in the interval $\syn\in [0,W_{max}]$, leading to $\langle W\rangle=W_{max}/2$. Therefore in order to modify $\langle W\rangle$ it is sufficient to change $W_{max}$. The model can be implemented on any network topology. Since it is difficult to directly measure the morphological synaptic connectivity in cortical systems, numerical studies have often implemented the structure of functional networks, as measured experimentally \cite{pajevic2009, massobrio2015, shefi2002morphological}. Therefore, the presented results are obtained on a three dimensional scale-free network where the degree distribution follows a power-law decay, $P(k)\propto k^{-\alpha}$ with $\alpha=2$\cite{chialvoScalefreeBrain}. We assign to each neuron an outgoing degree $k_{out}\in[2,100]$, which is then connected to other neurons with a distance dependent probability $P(r)\propto e^{-r/r_0}$. A configuration with an example avalanche is shown in Fig. \ref{fig:exampleConfig}.\\

Plastic adaptation governs how the structure of the network evolves and is a form of long-term plasticity. It follows the principles of Hebbian plasticity, which strengthens the synapses between neurons whose activity is correlated and weakens inactive synapses\cite{hebb}.  Due to short-term plasticity the synaptic weight $w_{ij}$ is a constantly changing quantity and is unsuitable to describe long-term plasticity. Therefore, in order to implement long-term plasticity $\syn$ is modified. More precisely, each time a synapse causes a firing event in a post-synaptic neuron, $\syn$ is increased proportionally to the potential variation in the post-synaptic neuron $j$,

\begin{equation}
\label{eq:LTP1}
\syn(t+1)=\syn(t)+\delta \syn(t),
\end{equation}
\begin{equation}
\label{eq:LTP2}
\delta \syn(t)=\alpha(v_j(t+1)-v_j(t)),
\end{equation}

\noindent where $\alpha$ is a parameter which determines the strength of plastic adaptation. It represents all possible biomolecular mechanisms which might influence the long-term adaptation of synapses. After an avalanche comes to an end, all $\syn$ are reduced by the average increase in synaptic strength\\
\begin{equation}
\Delta W=\frac{1}{N_s}\sum{\delta \syn},
\end{equation}\\

\noindent where $N_s$ is the total number of connections in the system. If a synapse is not used repeatedly it will progressively weaken and if its strength decreases below a minimal value, $\syn<10^{-4}$, the synapse is pruned. The weakening of synapses is a form of long-term depression. The combination of this form of activity dependent plasticity with the refractory time has been shown to be important in determining the structure of a neuronal network since it leads to the removal of loops\cite{meanfieldPaper,kozloski}. For the following simulations, plastic adaptation sculpts the long-term synaptic strengths of the initial scale-free network until the first synapse is pruned to not alter the scale-free degree distribution. After reaching this state long-term plasticity is suspended and new stimulations are then applied to measure the avalanche statistics.\\

\begin{figure}[htb]
    \centering
    \includegraphics[width=\columnwidth]{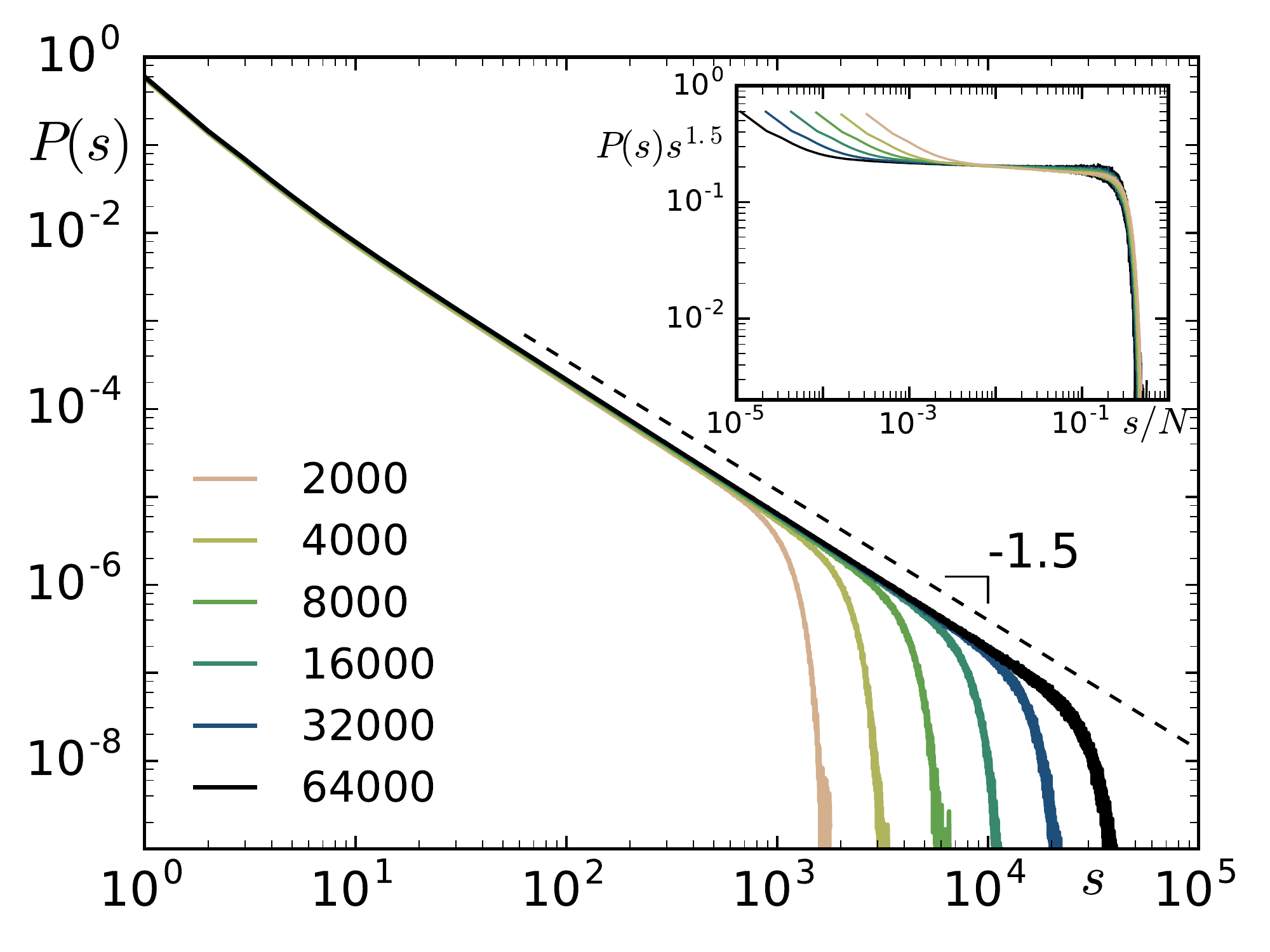}
    \caption{Avalanche size distribution on a scale free network. The distributions are obtained with $p_{in}=0.2$ and $t_r=1$. The values of $\langle W\rangle$ for each system size changes as shown in Fig. \ref{fig:controlParamScaling}. The inset shows a data collapse of the rescaled distributions.}
    \label{fig:controlParam}
\end{figure}

\begin{figure}[htb]
    \centering
    \includegraphics[width=\columnwidth]{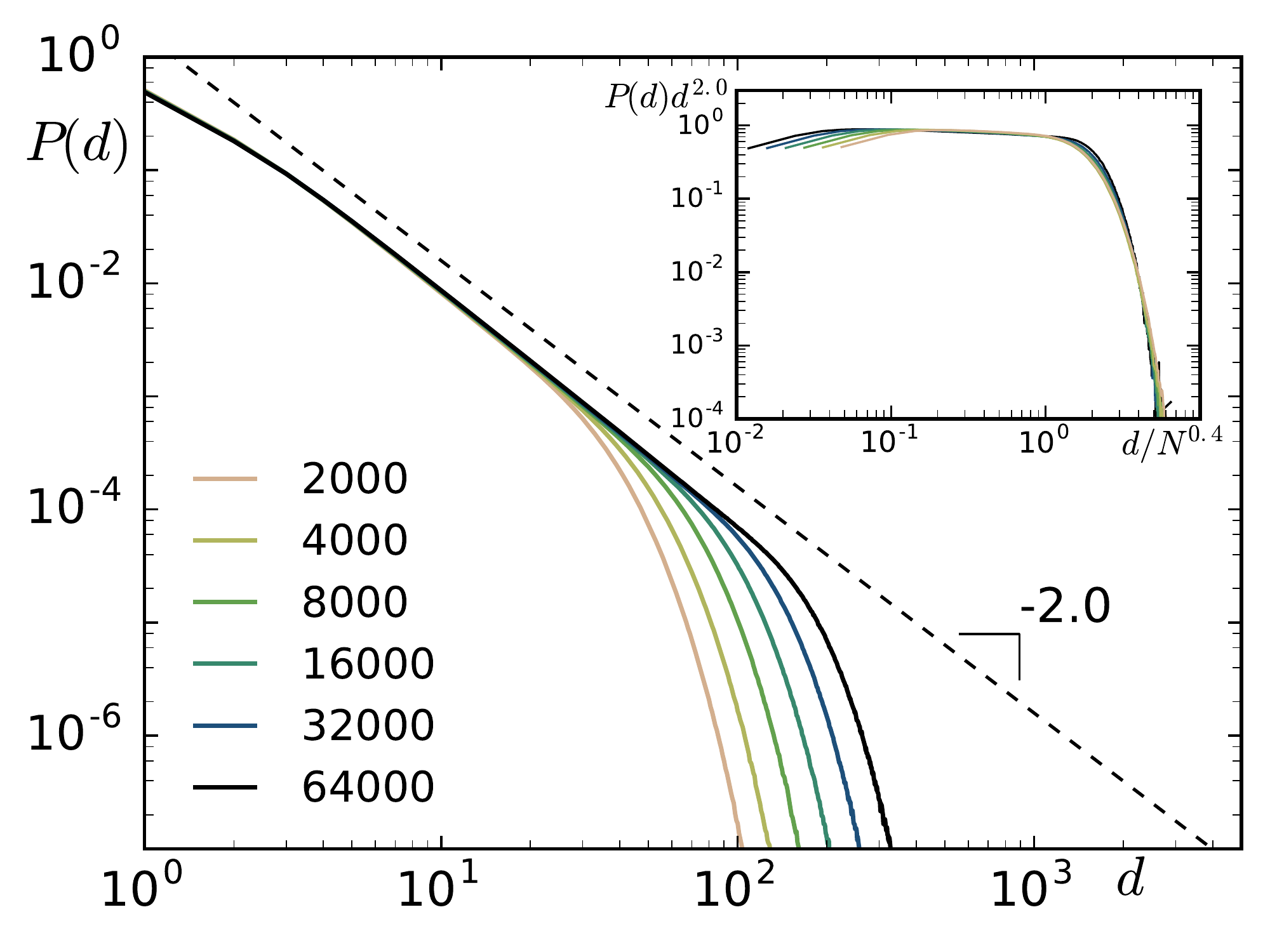}
    \caption{Avalanche duration distribution on a scale free network. The distributions shown are obtained with $p_{in}=0.2$ and $t_r=1$. The values of $\langle W\rangle$ for each curve changes with the system size (see Fig. \ref{fig:controlParamScaling}). The inset shows a data collapse of the rescaled distributions.}
    \label{fig:controlParamDuration}
\end{figure}

\section{Results}

\begin{figure}[htb]
    \centering
    \includegraphics[width=\columnwidth]{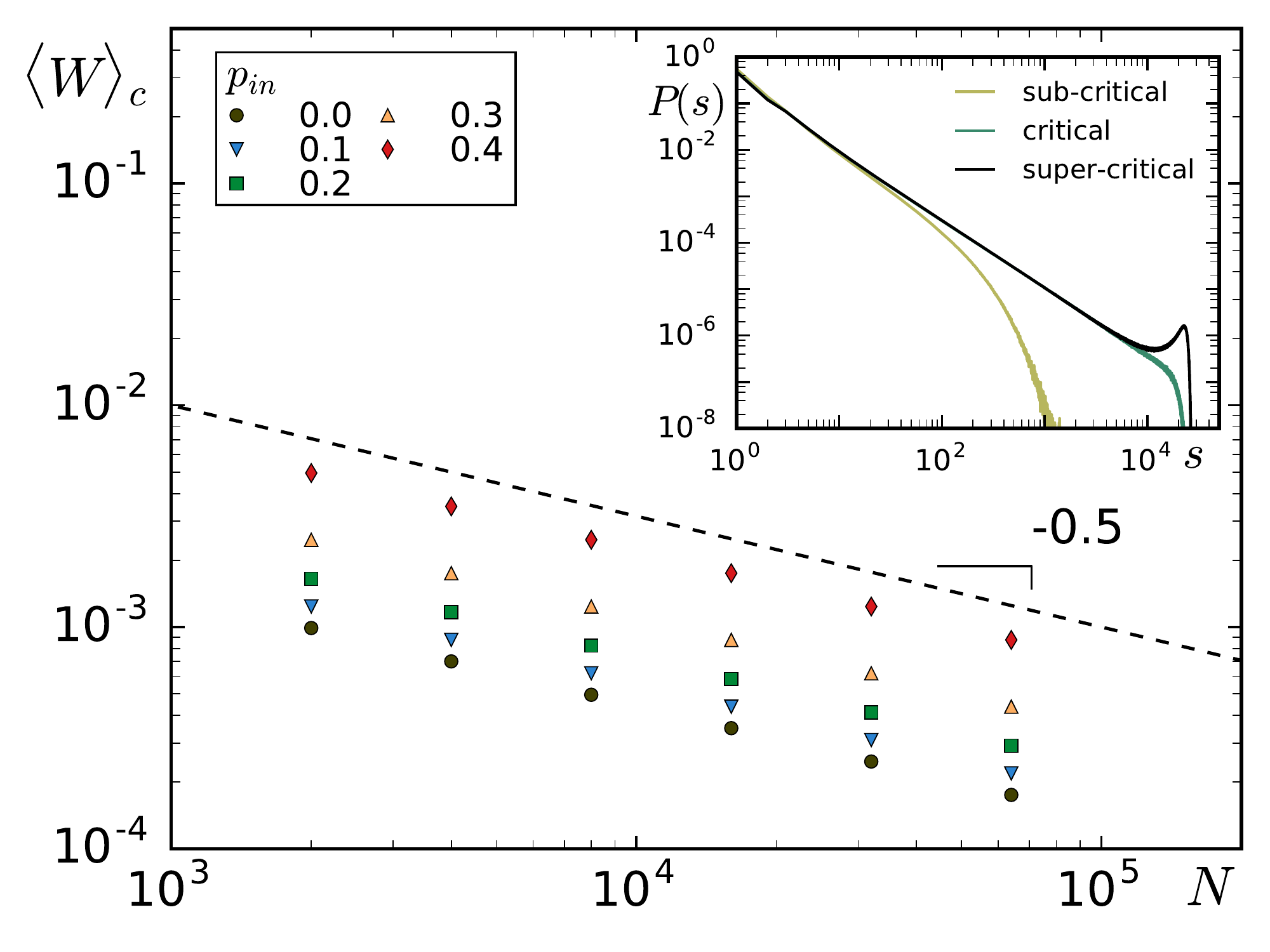}
    \caption{$\langle W\rangle_c$ scales as $\langle W\rangle _c\propto N^{-1/2}$ in order to have a cut-off scaling with the system size (see Fig. \ref{fig:controlParam}). This is observed for different values of $p_{in}$. The inset shows how changing the control parameter can lead to sub-, super- or critical behaviour for a network of 32000 neurons and $p_{in}=0.2$ with $\langle W\rangle=\{10^{-4},4*10^{-4},10^{-3}\}$.}
    \label{fig:controlParamScaling}
\end{figure}

Fig. \ref{fig:controlParam} shows the distribution of avalanche sizes with 20\% inhibitory neurons, $p_{in}=0.2$ and a refractory time of one time-step, $t_r=1$.  The distribution follows a power-law decay with an exponent of $1.5$ and an exponential cut-off. When the system size is increased the cut-off moves towards larger avalanches. The inset in Fig. \ref{fig:controlParam} shows the data collapse after rescaling the distributions, confirming that the cut-off scales $\propto N$ as suggested by the scale-invariant behaviour of the system. The corresponding duration distributions are shown in Fig. \ref{fig:controlParamDuration} and follow a power-law decay with exponent 2.0. The average long-term synaptic strength $\langle W\rangle$ is the control parameter which determines whether the network exhibits sub-, super- or critical behaviour (see inset Fig. \ref{fig:controlParamScaling}). We find that as the system size increases $\langle W\rangle$ needs to decrease proportional to $\langle W\rangle _c\propto N^{-1/2}$ to maintain a scaling cut-off. That the critical value of the control parameter changes with the system size has also been observed in Ref. \cite{levina}. Interestingly as the percentage of inhibitory neurons increases, an increase in $\langle W\rangle$ is sufficient to maintain criticality and the scaling with system size remains $\langle W\rangle _c\propto N^{-1/2}$, independent of $p_{in}$.
This scaling behaviour can be motivated by considering that as the system size increases so does the number of synapses, $N_s\propto N$. Each synapse recovers over time according to $\langle W\rangle$ and the total amount of neurotransmitter recovered $R_{tot}$ in the entire system is proportional to the number of synapses and therefore, $R_{tot}\propto N\langle W\rangle$. On the other hand, the dissipation of neurotransmitter $D$ is proportional to the average avalanche size $\langle s\rangle$. Larger avalanches have more firing events and according to equation (\ref{eq:dynSyn}) each firing event dissipates neurotransmitter. If the avalanche size distribution follows a power-law with exponent $\alpha$ than the average avalanche size scales as $\langle s\rangle=\int_{s_{min}}^N \! s^{-\alpha+1} \, \mathrm{d}x\propto N^{-\alpha+2}$ \cite{newman2005power}. With $\alpha=1.5$ one finds that the total amount of dissipation scales $D\propto N^{1/2}$. The ratio $R_{tot}/D=N^{1/2}\langle W\rangle$ determines the energy balance of the system. Therefore to maintain the energy balance as $N$ increases, $\langle W\rangle$ needs to scale as $\langle W\rangle\propto N^{-1/2}$.\\

It is important to note that the resulting model is not self-organising due to the presence of the control parameter $\langle W\rangle$. The critical value for the control parameter tends to zero in the thermodynamic limit, $N\rightarrow\infty$. In finite size systems $\langle W\rangle$ needs to be carefully tuned.

\section{Power spectrum}
The power spectral density (PSD) analysis is frequently employed in the investigation of neuronal activity to characterize both oscillations and temporal correlations\cite{lombardi2017chaos}. Magnetoencephalographic (MEG) and electroencephalographic (EEG) measurements, as well as LFPs of ongoing cortical activity have reported PSDs that decay with a power law $1/f^\beta$ \cite{novikov1997scale,bedard2006does, dehghani2010comparative,pritchard1992brain,zarahn1997empirical,linkenkaer2001long}. Measurements of human eyes-closed and eyes-open resting EEG yielded exponents of $\beta=1.32$ and $\beta=1.27$\cite{pritchard1992brain}. The measurements of $\beta$ varied over different brain regions and the standard error of estimate was in the interval [0.26,0.28]. A similar exponent, $\beta=1.33\pm0.19$ has been measured by \textit{Dehghani et al.}\cite{dehghani2010comparative}, who also reported varying exponents across different brain regions between $\beta=1$ and $\beta=2$. \textit{Novikov et al.} analyzed the power spectrum of MEG measurements and reported exponents of 0.98 for one subject and 1.28 for another\cite{novikov1997scale}. To summarize, many experiments report temporal correlations and measurements of the PSD frequently report a power-law decay $1/f^\beta$, with exponents in the interval [0.8,1.5] \cite{lombardi2017chaos}. Different exponents have been found in neural activity of epileptic patients. In the awake state $\beta$ was measured in the range [2.2,2.44] and in the slow wave sleep $\beta$ was in the range[1.6,2.87]\cite{he2010temporal}. In the following we analyze the PSD obtained from the presented neuronal model. The power spectrum of a temporal sequence is given by, $S(f)=\hat{a}(f)\hat{a}^*(f)$ where $\hat{a}(f)$ is the discrete Fourier transform of $a(t)$

$$\hat{a}(f)=\sum^{T-1}_{t=0} a(t)e^{-2i\pi kt/T}.$$

To measure the power spectrum of the activity in the model, the sequence $a_1(t)$ is recorded, where $a_1(t)$ is the total number of firing neurons at time $t$. During quiescent times neurons are stimulated by adding $\delta v=0.1$ to a random neuron. Each  stimulation counts as one time-step. Additionally we study another sequence defined as

\begin{equation}
\label{eq:sequence2}
a_2(t)=\left\{
\begin{array}{@{}rl@{}}
1,& \text{if } a_1(t)\geq 1\\
0,& \text{if } a_1(t)=0.\\
\end{array}
\right .
\end{equation}

\begin{figure}[htb]
    \centering
    \includegraphics[width=\columnwidth]{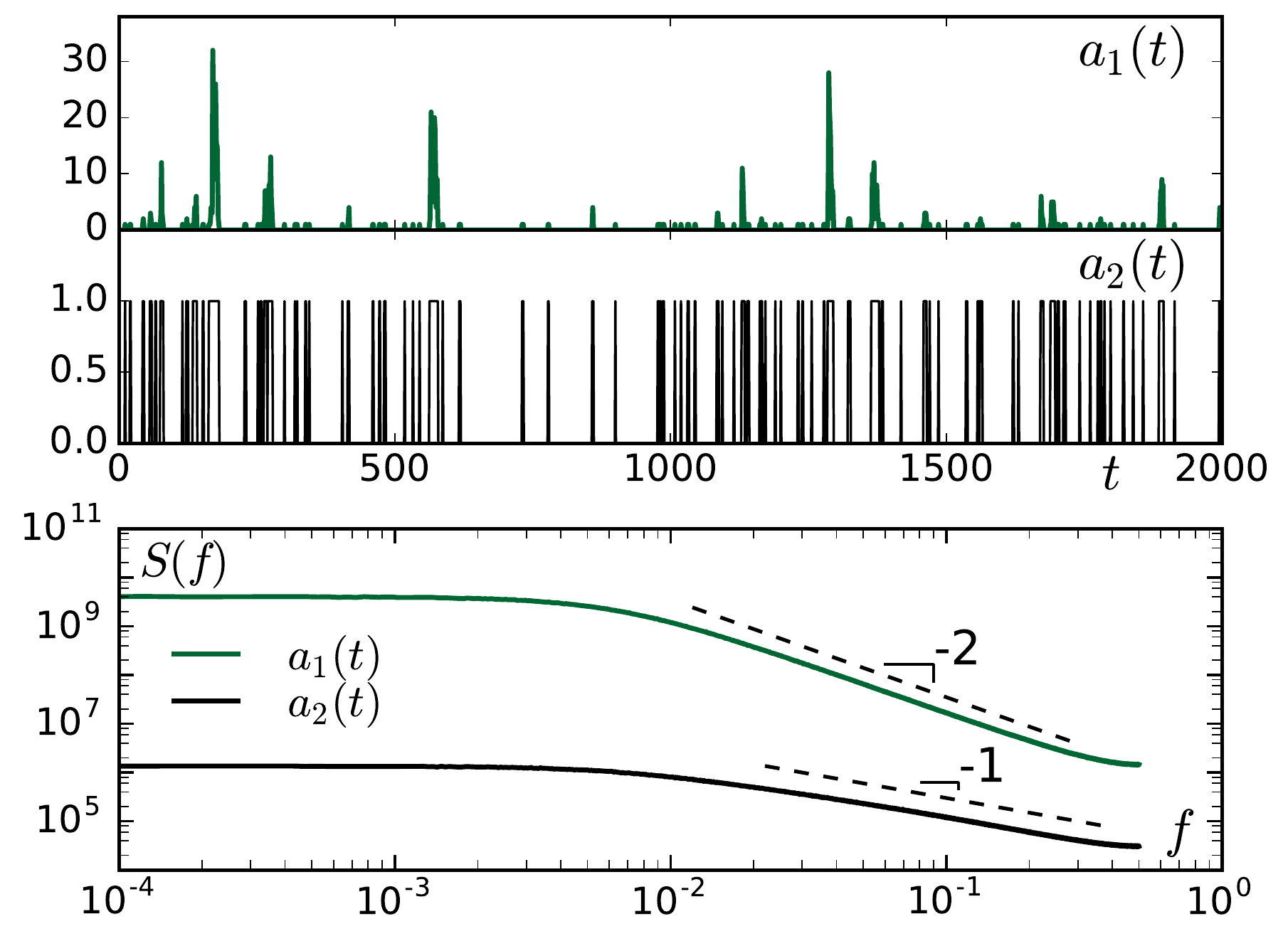}
    \caption{The upper part shows activity sequences $a_1(t)$ and $a_2(t)$. Below the corresponding power spectra are shown. The dashed lines show power laws with exponents $\beta=2$ and $\beta=1$ respectively.}
    \label{fig:powerspectrum}
\end{figure}

\begin{figure}[htb]
    \centering
    \includegraphics[width=\columnwidth]{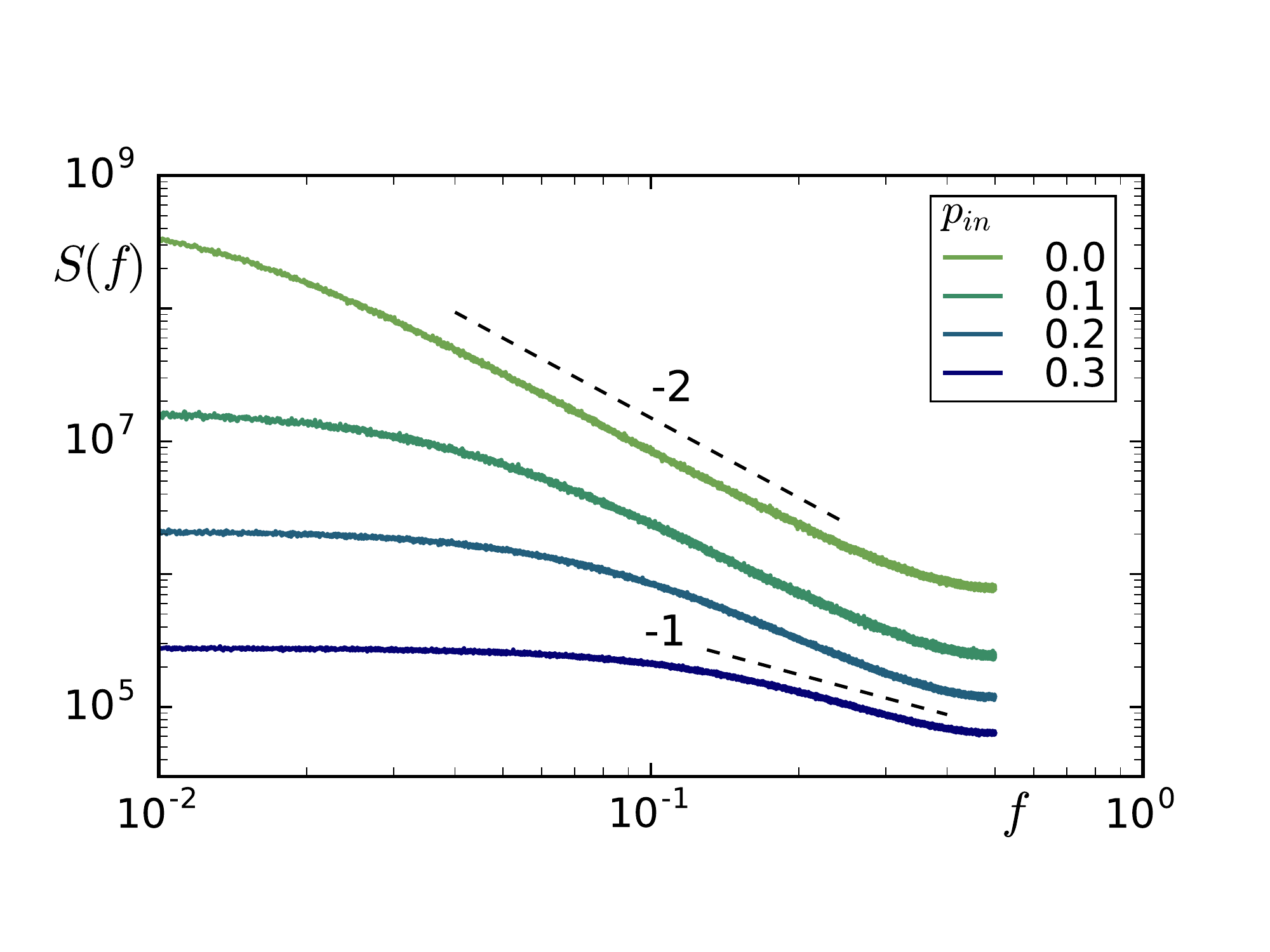}
    \caption{Increasing the percentage of inhibitory neurons while keeping $\langle W\rangle$ constant leads to a $1/f$ power spectrum for $p_{in}=0.3$ on a scale-free network of 16000 neurons, $t_r=1$.}
    \label{fig:pinPSD}
\end{figure}

An example of $a_1(t)$ and $a_2(t)$ is shown in  Fig. \ref{fig:powerspectrum}. These sequences are obtained for a network of 32000 neurons ($p_{in}=0.2$, $t_r=1$). The PSDs for the sequences $a_1(t)$ and $a_2(t)$ with $10^6$ time-steps are shown in the lower panel of Fig. \ref{fig:powerspectrum}. The power spectra differ significantly. For the sequence $a_1(t)$ it decays as $1/f^2$ before transitioning to white noise for low frequency. The $1/f^2$ power spectrum has also been observed in other critical systems such as the Bak-Tang-Wiesenfeld model \cite{jensen19891}. Conversely, $a_2(t)$ yields a PSD with a $1/f$ decay before it transitions to white noise. The crossover to white noise occurs, for both $a_1(t)$ and $a_2(t)$, approximately at the frequency corresponding to the cut-off of the avalanche duration distribution (see fig. \ref{fig:controlParamDuration}).
Another study of the power spectrum \cite{lombardi2017chaos} has shown that increasing the fraction of inhibitory neurons $p_{in}$ in a fully self-organised critical model leads to a PSD which decays with $1/f$ for the sequence $a_1(t)$. Therefore, the balance between excitation and inhibition has been considered as the origin of the $1/f$ power spectrum. The authors also report that the cut-off of the avalanche size distribution scales with $p_{in}$ and therefore avalanches as large as the system size become less probable for large $p_{in}$\cite{lombardi2017chaos}. In the model presented in this article these results are confirmed since an increase in inhibition (without adjusting $\langle W\rangle$) also leads to a $1/f$ power spectrum for sequence $a_1(t)$, see fig. \ref{fig:pinPSD}. In this case the amplitude of very large avalanches become smaller in size due to the dissipative role of inhibitory neurons and the system moves away from the critical point $\langle W\rangle_c$. These results suggest that avalanches with a size comparable to the system size affect the temporal correlations measured by the power spectrum. This is confirmed by the spectrum of series $a_2(t)$ in fig. \ref{fig:powerspectrum} in which case the avalanches all have the same amplitude over time and the temporal correlations are revealed, yielding a $1/f$ power spectrum.

\section{Learning}

\begin{figure}[htb]
    \centering
    \includegraphics[width=\columnwidth]{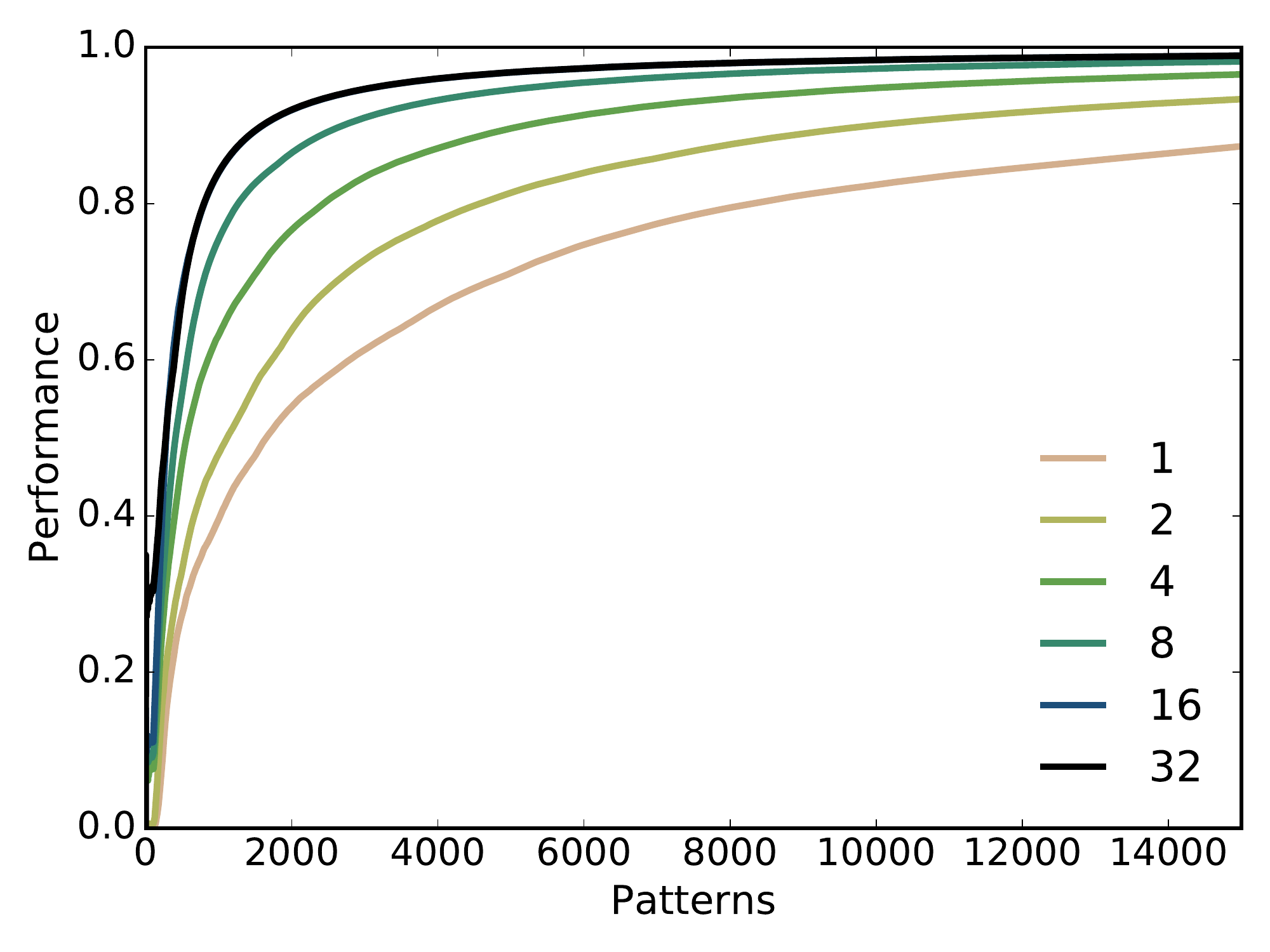}
    \caption{Learning performance versus the number of presented binary patterns of the XOR rule. The system consists of 3000 neurons with $p_{in}=0.2$, $t_r=1$. $\langle W\rangle$ is initialized to $\langle W\rangle=2*10^{-3}$. The different curves correspond to different numbers of input neurons.}
    \label{fig:xorPerformance}
\end{figure}

So far we have shown that the introduction of both long- and short-term plasticity is compatible with experimental data regarding avalanche statistics, branching ratio and power spectra. However the main role of long-term plasticity is its ability to sculpt the network over time to allow the system to learn. Here we study the systems ability to learn binary rules. The network can learn any binary rule but we focus here on the XOR rule as it is a non-linearly separable rule which makes the task more complex\cite{Kriesel2007NeuralNetworks}. The learning mechanism is inspired by previous work \cite{de2010learning,bak2001adaptive} and is based on the release of chemical signals, for example dopamine, which are known to mediate plasticity \cite{liu2000direct,reynolds2002dopamine,berger2017spatial}. The learning mechanism operates as follows. Two input neurons are defined at one side of the network which will provide the input to the system. On the other side of the network an output neuron is defined. All input and output neurons are assigned the maximal degree $k=100$ to ensure that they are well connected to the network (for the output neuron it is the incoming degree $k_{in}=100$). The two input neurons are triggered according to the possible binary combinations, i.e. one of the input neurons fires while the other remains inactive, $\{0,1\}$ and $\{1,0\}$, or both neurons fire at the same time $\{1,1\}$. The binary input where both neurons are inactive $\{0,0\}$ is omitted as it won't cause any activity. The response of the system is monitored at the output neuron. If the output neuron fires during an avalanche the response is $1$ otherwise it is $0$. The output neuron should therefore fire only if one of the input neurons is activated but not if both inputs fire simultaneously. To amplify the input signal additional input neurons can be defined which fire together. For example if four input neurons are defined the input $\{1,0\}$ triggers 4 neurons and $\{0,1\}$ triggers 4 other neurons. This still implements the XOR rule but amplifies the neuronal signal. As we will show, amplifying the input signal speeds up the convergence of the learning task. The learning scheme follows the principles of supervised learning and negative feedback\cite{bak2001adaptive}. The possible binary inputs are presented to the network in a random order and each time an avalanche is triggered. The activity propagates through the network and either activates the output neuron (1) or not (0). If the result is incorrect a feedback signal is released from the output neuron which modifies all synapses which participated in the previous avalanche, according to:

\begin{equation}
\label{eq:learningMech}
W_{ij}\rightarrow W_{ij}+\alpha Ef(d_{jo})
\end{equation}

\noindent where $E\in\{-1,0,1\}$ is the committed error. $E=-1$ signifies that the output neuron fired but should not have and $E=1$ means that the output neuron remained inactive but its activation was desired. Therefore depending on the error committed, synapses are either enhanced or weakened. $E=0$ corresponds to the correct output and consequently does not lead to any modification of synaptic strengths. $\alpha$ is the global learning strength and has the same meaning as in equation \ref{eq:LTP2}. Here $\alpha$ is set to $\alpha=0.3$ and modifying $\alpha$ does not qualitatively change the results presented here. $f(d_{jo})$ is a function of the distance $d_{jo}$ between the synapse and the output-neuron $o$. We consider the position of a synapse $ij$ to be the same as of the post-synaptic neuron $j$ to which it connects. For the function $f(d_{jo})$ we use an exponential decay, $f(d_{jo})=e^{-d_{jo}/d_0}$. The relevance of the spatial extent of this learning mechanism has been studied in detail in Ref. \cite{berger2017spatial}. How the learning rule modifies a synapse $j$, therefore, depends on the distance $d_{jo}$ from the source of the chemical signal, the plasticity rate $\alpha$ and the error $E$. We define the learning performance as $P=I_{correct}/I_{total}$, where $I_{total}$ is the total number of input signals presented to the network and $I_{correct}$ is the number of times the system gave the correct response. Fig. \ref{fig:xorPerformance} shows the performance of the system versus the number of binary patterns presented to the network. Results were obtained on a network of 3000 Neurons with $p_{in}=0.2$. The different curves correspond to different numbers of input neurons and results show that the network is able to learn XOR with a faster convergence when the input is amplified.

\section{Discussion}

The presented model includes several neurobiological ingredients such as both, short- and long-term plasticity, as well as the refractory time and inhibitory neurons. The introduction of short-term plasticity implies that the synaptic weights $w_{ij}$ are a constantly changing quantity. Therefore long-term plasticity cannot operate on the same synaptic weights as it would be dominated by the short-term fluctuations of $w_{ij}$. We therefore propose a long-term component of the synaptic strength, $\syn$, for each synapse. Stronger synapses have a larger $\syn$ which in turn will lead to a higher average short-term synaptic strength $w_{ij}$. Since critical dynamics relies on energy conservation an important question is how such an energy balance is maintained in neural systems. Synaptic interactions are of chemical nature and the individual firing events are non-conserving. This is especially true for inhibitory connections. Here we make use of dynamical synapses and show that increasing the amount of inhibition can be counteracted by increasing the long-term synaptic strengths. Furthermore, in the thermodynamic limit the tuning parameter tends to zero, $\langle W\rangle _c\propto N^{-1/2}$, independently of the percentage of inhibitory neurons. It has to be noted that the presence of a tuning parameter implies that the presented dynamics is not self-organising. However it is the tuning parameter which leads to scaling in the avalanche statistics even with large fractions of inhibitory neurons. The measured exponents are $1.5$ for the avalanche size and $2.0$ for the avalanche duration distribution, consistent with experimental studies\cite{plenz2014criticality,beggsPlenz2003}.\\

The spectral analysis of the neuronal activity yields a power spectrum decaying as $1/f^2$ for the sequence $a_1(t)$ which has also been observed in other critical systems\cite{jensen19891}. Conversely, the analysis of the sequence $a_2(t)$ results in a power spectrum with a $1/f$ decay. Recent results from a fully self-organised critical model \cite{lombardi2017chaos} show that an increase of inhibition leads to the $1/f$ power spectrum also in the sequence $a_1(t)$. We confirm these results and indeed increasing the fraction of inhibitory neurons $p_{in}$ while keeping $\langle W\rangle$ constant, leads to a decrease in the exponent of the power spectrum. As $p_{in}$ increases larger avalanches become less probable leading to a $1/f$ power spectrum for $p_{in}=0.3$, in accordance with several experimental studies\cite{novikov1997scale,bedard2006does, dehghani2010comparative, pritchard1992brain,zarahn1997empirical, linkenkaer2001long}. Finally we present results regarding the learning capabilities of the model. With distance dependent feedback signals the system is able to learn binary rules. We show the example of the XOR rule which is known to be difficult to learn as it is not linearly separable. The ability to learn XOR encourages the investigation of more complex tasks such as pattern recognition. This provides a foundation for future research of the effects of the combination of short- and long-term plasticity on the learning capabilities of neuronal systems.

\begin{acknowledgments}
The Authors would like to thank D. Berger for helpful discussions.
\end{acknowledgments}

% Create the reference section using BibTeX:
%\bibliography{ref.bib}

%merlin.mbs apsrev4-1.bst 2010-07-25 4.21a (PWD, AO, DPC) hacked
%Control: key (0)
%Control: author (8) initials jnrlst
%Control: editor formatted (1) identically to author
%Control: production of article title (-1) disabled
%Control: page (0) single
%Control: year (1) truncated
%Control: production of eprint (0) enabled
%

\end{document}